\documentstyle[aps,manuscript]{revtex}
\begin{document}
\draft 
\title{Photoassociative Frequency Shift in a Quantum Degenerate Gas}
\author{Jordan M. Gerton, Brian J. Frew, and Randall G. Hulet}
\address{Department of Physics and Astronomy and Rice Quantum Institute, Rice University,
    Houston, Texas 77251}
\date{\today}
\maketitle

\begin{abstract}
We observe a light-induced frequency shift in single-photon
photoassociative spectra of magnetically trapped, quantum
degenerate $^7$Li. The shift is a manifestation of the coupling
between the threshold continuum scattering states and discrete
bound levels in the excited-state molecular potential induced by
the photoassociation laser. The frequency shift is observed to be
linear in the laser intensity with a measured proportionality
constant that is in good agreement with theoretical predictions.
The frequency shift has important implications for a scheme to
alter the interactions between atoms in a Bose-Einstein condensate
using photoassociation resonances.
\end{abstract}
\vspace{0.25in}

Photoassociative spectroscopy of trapped atomic gases has been
used to measure interatomic interaction potentials with ultra-high
sensitivity \cite{Weiner99}. Although most photoassociation
experiments with trapped atoms have been performed at temperatures
near 1 mK, sub-$\mu$K temperatures can be achieved using
evaporative cooling.  This is the temperature regime that recent
experiments on Bose-Einstein condensation (BEC)
\cite{Anderson95,Bradley95,Davis95d} have been conducted. There
are several important improvements to photoassociative
spectroscopy that are realized with a quantum degenerate gas.
Since the energy spread of atoms with temperatures under 1 $\mu$K
is less than 20 kHz, spectroscopic precision can be increased to
unprecedented levels. Furthermore, it was recently pointed out
that the rate of photoassociation increases with increasing phase
space density, $n \lambda_D$, where $n$ is the atomic density and
$\lambda_D$ is the thermal de Broglie wavelength
\cite{Javanainen98}. Finally, at sub-$\mu$K temperatures the
spatial extent of the trapped gas can be very small, enabling
tighter focusing of the photoassociation laser beam and,
therefore, higher light intensities. We have exploited these
enhancements to investigate the effect of light intensity on
single-photon photoassociation spectra of a magnetically trapped,
evaporatively cooled gas of $^7$Li. In particular, we have
measured a spectral shift proportional to the light intensity
\cite{Javanainen98,Fedichev96,Bohn97,Bohn99}. This shift is
relevant to proposed schemes for utilizing photoassociation to
alter the interactions between atoms in a Bose-Einstein condensate
\cite{Fedichev96,Bohn97} and for producing ultracold, trapped
molecules
\cite{Javanainen98,Band95,Bohn96,Cote97,Vardi97,Fioretti98,Julienne98,Drummond98,Takekoshi98,Nikolov99,Javanainen99,Wynar00}.

The apparatus used in this experiment, which has been used to
produce BEC of $^7$Li, has been described previously
\cite{Sackett97b}. Permanent magnets establish an Ioffe-Pritchard
type trap with a depth of $10$ mK and a bias field of 1004 G at
the trap center. Approximately $5 \times 10^8$ atoms in the $F =
2$, $m_F = 2$ hyperfine sublevel of $^7$Li are directly loaded
into the trap from a laser-slowed atomic beam using
three-dimensional optical molasses. Following loading, the laser
beams are extinguished and the atoms are evaporatively cooled
using a microwave field to selectively spin-flip, and thereby
remove the hottest atoms.  The final temperatures are between 400
and 650 nK, corresponding to between $3\times10^5$ and
$1\times10^6$ atoms. Under these conditions, the gas is quantum
degenerate, although the fraction of atoms in the condensate is
small due to attractive interactions in lithium \cite{Bradley97a}.

Following evaporation, a photoassociation laser beam of frequency
$\omega_1$, and intensity $I$, is passed through the trapped
atoms. A schematic representation of the relevant molecular
potentials and energy levels is shown in Fig.~\ref{fig:pa}. A
vibrationally and electronically excited molecule may form when
$\omega_1$ is tuned to resonance between the continuum level of
two colliding atoms and a bound level in the excited-state
molecular potential. The excited molecule may spontaneously decay,
most probably into a pair of hot atoms or possibly into a
ground-state molecule, resulting in a detectable reduction of
trapped atoms. The trapped cloud is probed {\it in situ} using the
phase contrast imaging technique described in
Ref.~\cite{Sackett97b} in order to determine the number of
remaining atoms $N$. Since photoassociation removes a significant
number of atoms, only one image may be obtained per evaporative
cooling cycle. Therefore, to build up a resonance curve, the
entire cycle is repeated many times for different values of
$\omega_1$.

The photoassociation light is derived from a low-power ($\sim$10
mW), grating-stabilized,  external cavity diode laser. The laser
is side-locked to a 3 GHz free-spectral-range (FSR) scannable
Fabry-Perot cavity in order to reduce acoustical jitter to an RMS
amplitude of $\sim$1 MHz as measured with an optical spectrum
analyzer. A 750 MHz FSR scanning etalon is used to measure the
relative frequency separation between the diode laser and another
laser which is locked to the $2S_{1/2} - 2P_{3/2}$ atomic
resonance. Slow feedback to the 3 GHz cavity maintains this
frequency separation to within $\sim$3 MHz. An $\sim$8 mW beam
from the diode laser is injected into a tapered optical amplifier,
providing up to 300 mW of output power at the injected frequency.
An acousto-optic modulator and a mechanical shutter are used to
chop the amplified beam on and off. The beam is subsequently
coupled into a single-mode optical fiber, reducing pointing jitter
and intensity variation across the beam profile. The output of the
fiber, limited to 70 mW, is focussed at the position of the atoms
to a $1/e^2$ intensity radius that ranges between 60 and 120
$\mu$m. The laser beam waist is always larger than the $1/e$
density radius of the atoms of $\sim$40 $\mu$m.  The
photoassociation laser beam is directed nearly parallel to the
laser beam used for imaging the atom cloud, which allows for the
imaging optics to be used to ensure spatial overlap of the
photoassociation laser beam with the atom cloud.

The laser frequency $\omega_1$ is tuned to near resonance with the
$v=69$ vibrational level of the excited molecular potential. Since
the binding energy of this level, 854 GHz, is much larger than
either the 10 GHz fine-structure splitting of the $2P$ atomic
state or any hyperfine interaction, the total electronic spin
${\bf S}$ decouples from both the electronic orbital angular
momentum ${\bf L}$ and the total nuclear spin ${\bf I}$.  In this
case, the light field only couples to ${\bf L}$, and ${\bf S}$ and
${\bf I}$ do not change during the photoassociation transition.
Therefore, the selection rules for the total spin ${\bf G} = {\bf
S} + {\bf I}$, and its projection $M_G$, are $\Delta {\bf G} = 0$
and $\Delta M_G = 0$ \cite{Abraham96}, and there is no first-order
Zeeman shift in the transition energy. Since both the ground state
and excited molecular potentials are $\Sigma$ states,
corresponding to zero projection of ${\bf L}$ onto the
internuclear axis, there is no change in the projection of ${\bf
L}$, and the transition dipole is oriented along the trap magnetic
field bias direction (0,0,1). Geometric constraints, however,
require the photoassociation laser beam to propagate along the
(1,1,1) direction, so that only $2/3$ of the light intensity can
be polarized along the transition dipole.  In this paper, the
reported intensities correspond to those actually measured.

Figure~\ref{fig:res} shows several resonance curves corresponding
to different values of $I$. The data points are normalized to
background images obtained without the photoassociation pulse to
account for drift in trap loading conditions. The data clearly
demonstrate that the resonance is red-shifted with increasing $I$.
As the intensity is varied the photoassociation pulse time is
adjusted to maintain a relatively large and constant signal size.
The resonance spectral widths are observed to be between 20 and 30
MHz. The natural linewidth for a long-range vibrational level,
such as $v = 69$, is $\sim$2$\Gamma_a$ \cite{Thorsheim87}, where
$\Gamma_a = 2 \pi \times 5.9$ MHz is the natural linewidth of the
atomic $2P$ excited state of $^7$Li. There are several other
mechanisms which contribute to the broadening of the observed
lineshapes. The relatively large depth of signal leads to a
saturation broadening, which is the case even for low $I$ since
the pulse duration is extended to maintain a relatively constant
signal size. Additionally, inhomogeneous broadening caused by the
variation of laser intensity across the spatial extent of the atom
cloud is expected to contribute as much as 10 MHz to the width for
the largest values of $I$.  The temperature of the gas is
sufficiently low that thermal broadening is negligible compared
with $\Gamma_a$. The observed resonance curves are sufficiently
symmetrical to be fit reasonably well to a Lorentzian lineshape in
order to locate the resonance center. Figure~\ref{fig:intens}
shows the frequency shift of the measured resonance peaks as a
function of $I$. These data fit a line of slope (-245 $\pm$ 10)
kHz-cm$^2$/W.

The origin of the frequency redshift is explained in
Ref.~\cite{Bohn99}. It arises from coupling the various threshold
scattering states to the excited-state bound level $v$ and will be
a generic feature of near-threshold resonant scattering. In
particular, in the Fano theory for a bound state coupled to the
continuum, the energy shift is proportional to the integral
\cite{Fano61}
\begin{equation}\label{eqn:fano}
  \delta \epsilon \propto \int dE^{\prime} \frac{V(E^{\prime})^2}{E -
  E^{\prime}} \ ,
\end{equation}
where $V(E^{\prime})$ represents the density of continuum states
at energy $E^{\prime}$.  The integrand in Eqn.~\ref{eqn:fano} is
positive when $E^{\prime} < E$ and negative when $E^{\prime} > E$,
where $E$ is the unperturbed resonance energy. The density of
continuum states $V(E^{\prime})$ always increases versus energy
$E^{\prime}$ no matter if the gas is confined to a trap, or in
free space. Therefore, the negative part of the integral will
contribute more strongly, and Eqn.~\ref{eqn:fano} gives rise to a
red shift.

References \cite{Fedichev96} and \cite{Bohn99} give simple,
approximate expressions for the frequency shift.  In
Ref.~\cite{Fedichev96}, the magnitude of the frequency shift is
given as $\beta \Omega^2 / \delta \varepsilon_v$, where, in their
notation, $\Omega^2 = (\Gamma_a^2 / 2) \, I / I_{sat}$, $I_{sat} =
5.1$ mW/cm$^2$ is the saturation intensity of the atomic
resonance, $\delta \varepsilon_v$ is the energy difference between
vibrational levels $v$ and $v+1$, and $\beta$ is a numerical
factor which characterizes the overlap integral between the ground
and excited-state wave functions. As per the discussion following
Eq.~(7) of Ref.~\cite{Fedichev96}, $\beta \approx 0.8 \pi^2 (1 -
a_T/r_t)$ in the low temperature limit, where $a_T = - 27.6~a_0$
is the ground-state triplet scattering length \cite{Abraham97},
$a_0$ is the Bohr radius, and $r_t$ is the classical outer turning
point of vibrational level $v$. For the $v = 69$ level, $r_t =
44~a_0$ and $\delta \varepsilon = 120.6$ GHz \cite{Abraham95b},
giving 242 kHz-cm$^2$/W for the magnitude of the theoretical
light-induced shift, when the factor of $2/3$ accounting for the
polarization of the photoassociation beam is taken into account. A
prediction of 250 kHz-cm$^2$/W was obtained from the authors of
Ref.~\cite{Bohn99}. Both of these predictions agree quite well
with the data.

The primary subject of Ref.~\cite{Fedichev96} is not the
light-induced frequency shift, but rather an optical method for
altering the interactions between ground-state atoms using
photoassociation resonances.  If the laser frequency is tuned near
a molecular resonance, there is some probability that the atom
pair will be excited to the upper potential shown in
Fig.~\ref{fig:pa}. If the laser detuning is made sufficiently
large, the probability of excitation is small, but since the
interaction potential of the excited state is much deeper than
that of the ground state in the long-range region, a small
excitation amplitude can have a substantial effect on the overall
interaction \cite{Fedichev96,Bohn97}. The degree of change in
interaction induced by the photoassociation laser, as well as the
amount of induced loss, depends sensitively on the frequency
detuning of the laser from the molecular resonance, necessitating
a detailed understanding of the light-induced frequency shift
described above.

This scheme for altering the interactions between atoms is an
optical analog to magnetically-tuned Feshbach resonances
\cite{Tiesinga93}, which have been used to alter the interactions
between atoms in Bose-Einstein condensates
\cite{Inouye98,Cornish00}. Evidence for these optically-induced
Feshbach resonances has recently been obtained in a gas of
ultracold sodium atoms \cite{Fatemi00}. There are several possible
advantages of an optical method, such as the speed in which
optical fields may be switched, the generality of the method for
different atomic species, and the simplicity of applying optical
fields, in comparison to changing magnetic field strengths. Such a
technique may be particularly useful for probing the dynamics of
condensate collapse in the case of attractive interactions
\cite{Sackett99,Gerton00,Roberts01}, and we have performed a
detailed analysis for a condensate of $^7$Li atoms.

Spontaneous light scattering induced by the photoassociation
laser, even when detuned far from the molecular resonance,
severely limits the utility of this technique for altering the
interaction between condensate atoms. Our analysis suggests that
this technique cannot be used to produce large condensates of
$^7$Li atoms by making the self-interactions repulsive, because
the induced loss rate exceeds the condensate growth rate. On the
other hand, the interaction can be made more attractive for long
enough ($\sim$10 ms) to induce a collapse for relatively small
condensates without significant spontaneous scattering losses.
Using $v=69$, we calculate that the magnitude of the scattering
length $a$ can be made three times larger for $I = 600$ W/cm$^2$
and $\Delta = -165$ MHz.  To achieve such an intensity with the
$\sim$70 mW of available laser power, the Gaussian $1/e^2$
intensity radius of the laser beam must be set to $\sim$90 $\mu$m.
Under these conditions, however, we observe that the laser beam
produces a substantial loss of trapped atoms accompanied by a
significant change in shape of the atom cloud. We attribute these
observations to heating of the atoms caused by the strong dipole
force which arises from the large intensity gradient across the
atom cloud.  We conclude that substantially more laser power is
required to make even modest changes to the scattering length
using this technique.

In conclusion, we find that the light-induced shift is
substantial, and must be taken into account, for example, in
schemes for producing trapped ultracold molecules using
photoassociation
\cite{Javanainen98,Band95,Bohn96,Cote97,Vardi97,Fioretti98,Julienne98,Drummond98,Takekoshi98,Nikolov99,Javanainen99,Wynar00}.
It is important to understand the laser-induced frequency shift
since the success of these molecule production schemes depends on
detailed knowledge of the intermediate-state spectral location.
The origin of the shift is well understood in terms of continuum
coupling, and its magnitude, predicted by semi-analytic theories,
is in excellent agreement with experiment.

The authors thank Ionut Prodan for help with the analysis.  This
work was partially funded by grants from the National Science
Foundation, the Office of Naval Research, the National Aeronautics
and Space Administration, and the Welch Foundation.


\begin{figure}
\caption{Relevant molecular potentials and energy levels. The
upper potential dissociates to a $2S_{1/2}$ and a $2P_{1/2}$ atom
while the lower potential dissociates to two $2S_{1/2}$ atoms. The
spontaneous emission rates $\gamma_b$ and $\gamma_f$ are to bound
and free states, respectively. The binding energy of the $v = 69$
level is given relative to dissociation.  A laser of frequency
$\omega_1$ drives the transition between a state of two free atoms
and an electronically excited molecule.  Since the atoms are
spin-polarized, they only couple to the spin-triplet potentials.}
\label{fig:pa}
\end{figure}

\begin{figure}
\caption{Single-photon photoassociation resonance curves for the
$v = 69$ vibrational level of the $1^3 \Sigma^+_g$ potential of
$^7$Li$_2$. The signal is the normalized number of atoms remaining
following the photoassociation pulse. The intensities, Gaussian
$1/e^2$ intensity radii, and pulse durations are: 290 W/cm$^2$,
120 $\mu$m, and 140 $\mu$s (solid squares); 53 W/cm$^2$, 60
$\mu$m, and 1 ms (open circles); and 12 W/cm$^2$, 95 $\mu$m, and
10 ms (solid triangles). At higher intensity, the signal baselines
do not normalize to unity as the large intensity variation across
the spatial extent of the atoms leads to a large dipole force
perturbation.} \label{fig:res}
\end{figure}

\begin{figure}
\caption{Photoassociation frequency shift as a function of light
intensity. The points correspond to the peaks of measured spectra
obtained as described in the text. The line is a fit to these
points giving a slope of (-245 $\pm$ 10) kHz-cm$^2$/W. The error
bars indicate the uncertainty in the fitted line center. The
uncertainty in the light intensity is $\pm 10\%$.  The quoted
uncertainty in the fitted slope is the change required to increase
the value of the goodness-of-fit parameter (unreduced $\chi^2$) by
unity.} \label{fig:intens}
\end{figure}

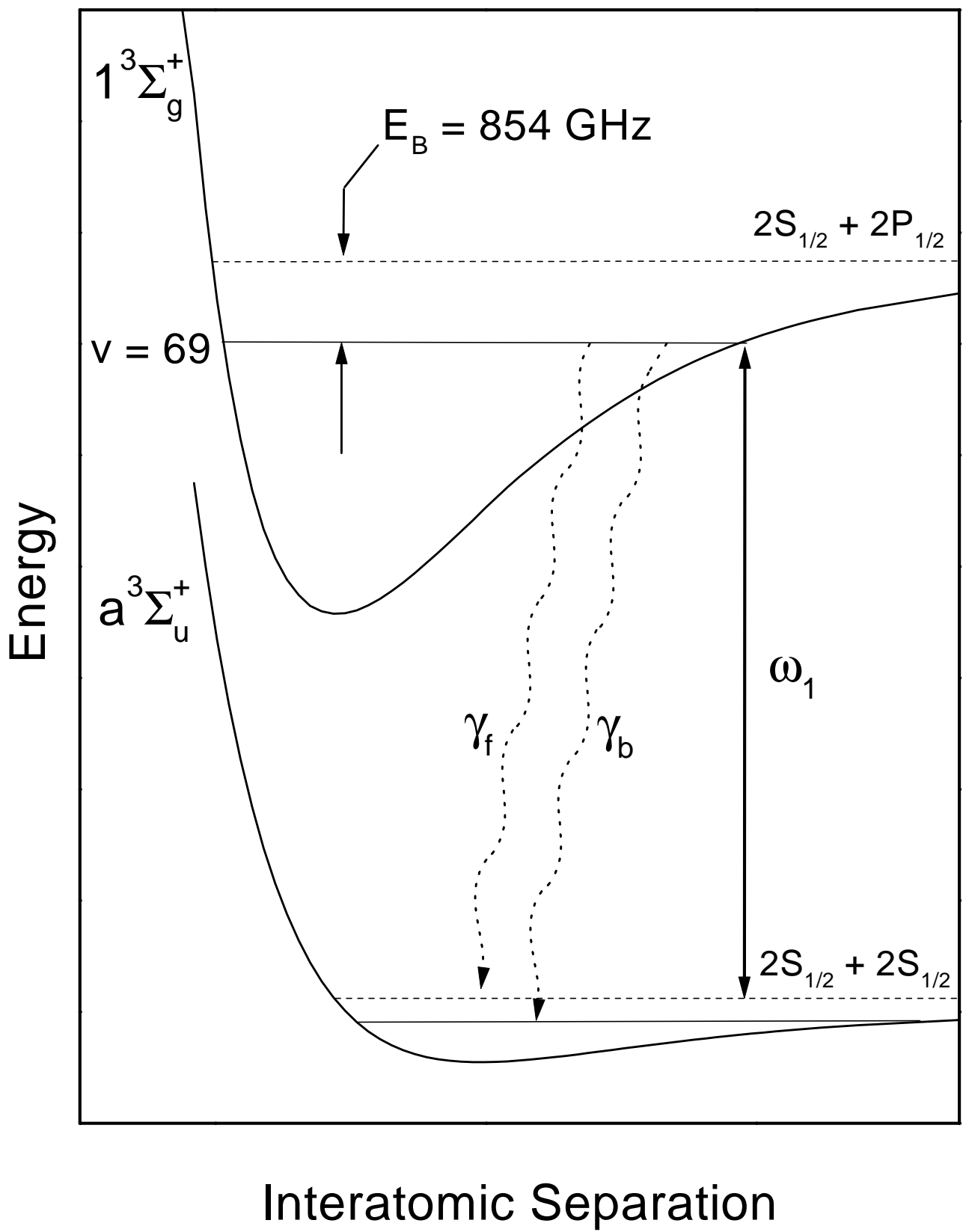
\end{document}